\documentclass[conference]{IEEEtran}
\usepackage{amsmath, fancybox, epsfig, amssymb, color, latexsym, graphicx, accents, subfigure, bm}

\begin{document}
\title{A Non-Disjoint Group Shuffled Decoding for LDPC Codes}
\author{\authorblockN{Yen-Cheng Hsu, Tofar C.-Y. Chang, Yu T. Su, and Jian-Jia Weng}
        \authorblockA{Institute of Communications Engineering \\National Chiao Tung University\\
        Hsinchu, 30010, TAIWAN\\
        Emails: \{ybressap.cm99g, tofar.cm96g\}@nctu.edu.tw,~ytsu@mail.nctu.edu.tw,~jianjiaweng@gmail.com~}
}
\maketitle
\begin{abstract}
To reduce the implementation complexity of a belief propagation
(BP) based low-density parity-check (LDPC) decoder, shuffled BP
decoding schedules, which serialize the decoding process by
dividing a complete parallel message-passing iteration into a
sequence of sub-iterations, have been proposed. The so-called
group horizontal shuffled BP algorithm partitions the check nodes
of the code graph into groups to perform group-by-group
message-passing decoding. This paper proposes a new grouping
technique to accelerate the message-passing rate. Performance of
the proposed algorithm is analyzed by a Gaussian approximation
approach. Both analysis and numerical experiments verify that the
new algorithm does yield a convergence rate faster than that of
existing conventional or group shuffled BP decoder with the same
computing complexity constraint.
\end{abstract}

\section{Introduction}
Low-density parity-check (LDPC) codes with belief propagation (BP)
or so-called sum-product algorithm (SPA) based decoder can offer
near-capacity performance. The SPA decoder, however, suffers from
low convergence rate and high implementation complexity. To
improve the rate of convergence and reduce implementation cost,
serialized BP decoding algorithms which partition either the
variable nodes (VNs) \cite{JZhang'05} or the check nodes (CNs)
\cite{DVBS2'06} of the corresponding bipartite graph into multiple
groups were introduced. These two classes of serial SPA algorithms
are called vertical and horizontal group shuffled BP decoding
algorithms, respectively. More recent related works can be found
in \cite{Sharon'07} -\cite{Yang'10}. These practical alternatives
use serial-parallel decoding schedules that perform sequential
group-wise message-passings and have the advantage of obtaining
more reliable extrinsic messages for subsequent decoding within an
iteration.

In this paper, we focus on the horizontal group shuffled BP
decoding algorithms as they provide more advantages in hardware
implementation \cite{JZhang'05} \cite{Yang'10}. For the sake of
brevity, group shuffled BP (GSBP) stands for horizontal group
shuffled BP (HGSBP) throughout this paper. For conventional GSBP
schedules, the CNs are divided into a number of groups such that
each CN belongs to just one group. A decoding iteration consists
of several sub-iterations. Each sub-iteration updates in parallel
the log-likelihood ratios (LLR) associated with the VNs connecting
to the CNs in the same group. Hence within a sub-iteration,
message-passing is performed on the bipartite subgraph that
consists of the CNs of a group and all the VNs connecting to these
CNs. Unlike conventional group shuffled (GS) schedules which
partition either VNs or CNs into disjoint groups, we propose a GS
decoding schedule which divides CNs into non-disjoint CN groups.
Such a CN grouping results in larger connectivity of consecutive
subgraphs (CoCSG) associated with two neighboring CN groups, where
the CoCSG, denoted by $\ell$, refers to the the average number of
VNs connecting the CNs of, say, the $k$th group and the VNs which
are also linked to the CNs of the previous, i.e., $(k-1)$th, CN
group. A larger CoCSG means more information will be forwarded
from the previous sub-iteration and thus provides opportunities
for improved decoding performance. We demonstrate by using both
simulation and analysis that the proposed GSBP is indeed capable
of offering significant performance gain and additional
performance-complexity-decoding delay tradeoffs. Since our
division on the CNs yields CN groups with a nonempty intersection
for any two neighboring groups, we refer to the resulting decoding
schedule as non-disjoint group-shuffled belief propagation
(NDGSBP) in subsequent discourse.

To analyze the performance of iterative LDPC decoding algorithms
in binary-input additive white Gaussian noise (BI-AWGN) channels,
approaches such as density evolution (DE), Gaussian approximation
(GA), and extrinsic information transfer (EXIT) charts have been
proposed \cite{Rich.'01}-\cite{ZRP'10}. We adopt the GA approach
\cite{Chung'01} \cite{ZRP'10} as it requires just the tracking of
the first two moments which are sufficient to completely
characterize the probability densities. Moreover, if a consistency
condition is met \cite{ZRP'10}, we need to track only the means of
related likelihood parameters.

The rest of this paper is organized as follows. In Section
\ref{section:algorithms}, we explain the basic idea of the new
grouping method, provide relevant parameter definitions and
present the NDGSBP decoding algorithm. The corresponding GA-based
performance analysis is given in Section \ref{section:analysis}.
Section \ref{section:performance} provides numerical performance
examples of the our algorithm, estimated by both computer
simulations and analysis. Finally, concluding remarks are drawn in
Section \ref{section:conclusions}.

\section{Non-Disjoint Group Shuffled Belief Propagation
Algorithm}\label{section:algorithms}
\begin{figure}
\begin{center}
\epsfxsize=3.5in
\includegraphics[height=1.3in,width=2.6in,angle=0]{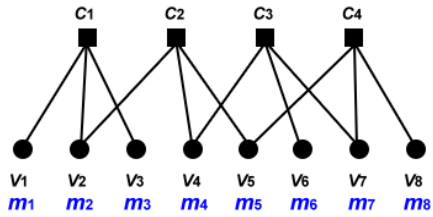}
\caption{\label{egfig}The Tanner Graph of A Linear Block Code.}
\end{center}
\end{figure}
\subsection{Why GS decoding with non-disjoint groups?}
Consider the decoding sub-iteration which performs VN-to-CN and
then CN-to-VN message passing for the CNs of the $k$th group and
all connecting VNs. If (at least) one of the VNs is linked to some
CNs in other (CN) groups which have been processed in the same
decoding iteration before (i.e., whose group indices are smaller
than $k$), then other connecting VNs which have no such links will
benefit from receiving more newly updated messages. We use a
simple linear code and its associating Tanner graph shown in Fig.
\ref{egfig}, where there are four CNs $\{c_1,c_2,c_3,c_4\}$ and
eight VNs $\{v_1,v_2,\dots,v_7,v_8\}$, to explain this effect. Let
the messages the VNs carry be denoted by $m_1,m_2,\dots,m_7,m_8$.
In a conventional BP decoding iteration, each VN receives the
messages from its neighboring VNs which are linked through some
VNs. For instance, $v_4$ and $v_6$ are updated by the messages
$\{m_2,m_5,m_6,m_7\}$ and $\{m_4,m_7\}$, respectively. For the
GSBP decoding with two CN groups $\{c_1,c_2\}$ and $\{c_3,c_4\}$,
$v_4$ receives $\{m_2,m_5\}$ in the first sub-iteration and
$\{m_2,m_5,m_6,m_7\}$ in the second sub-iteration while $v_6$ is
updated by $\{m_2,m_4,m_5,m_7\}$ in which $m_2$ and $m_5$ are the
messages forwarded by $v_4$ because of its connection to the
second CN group and will help improving the convergence.
Obviously, the amount of messages the CNs in the $k$ group receive
from VNs connected to CNs belonging to the $j$th group, $j < k$
depends on the code structure and the grouping of CNs. If we limit
our attention to the case $j=k-1$, the single parameter $\ell$
defined in the introductory section can be used to quantify the
average amount of messages received from the previous
sub-iteration and the grouping should try to maximize this number.


To simplify our systematic non-disjoint grouping method, we assume
identical group cardinality, $N_G$, and denote the number of CN
groups by $G$ so that $G \times N_G=M$ is the number of CNs. We
define the overlapping ratio $r$ as the ratio between the size of
the intersection between two neighboring CN groups and $G$. Then,
we have, $G N_G-(G-1)N_G r=M$.

We arbitrary select $N_G$ CNs to form the first CN group. The
$k$th ($k > 1$) group includes $r\cdot N_G$ CNs randomly chosen
from the $(k-1)$th group and $(1-r)\cdot N_G$ CNs from the CNs
which do not belong to any of the earlier groups. Therefore, a CN
does not necessarily belong to only one group anymore. As an
illustration, we consider the grouping $(r, G. N_G)=(0.5, 3,2)$ on
the Tanner graph of Fig. \ref{egfig} again. Let the first group be
$\{c_1, c_2\}$, the second one be $\{c_2, c_3\}$ and the third one
be $\{c_3, c_4\}$. In the first sub-iteration, $v_2$ and $v_4$
receive $\{m_1,m_3,m_4,m_5\}$ and $\{m_2,m_5\}$, respectively.
$v_4$ and $v_6$ receive $\{m_1,m_2,m_3,m_5,m_6,m_7\}$ and
$\{m_2,m_4, m_5,m_7\}$ in the second sub-iteration, in the final
sub-iteration, $v_6$ will be updated by $\{m_1,m_2,m_3,
m_4,m_5,m_7\}$. In short, for conventional BP, a VN can just
collect information from VNs which are two-edge away in one
iteration; for GSBP decoding, a VN has the opportunity to obtain
the messages from four-edge-apart VNs; and for the proposed NDGSBP
decoding algorithm, it is possible that a VN obtains the messages
from VNs which are more than six-edge away if we select the
overlapping ratio and CNs carefully. With fixed degree of
parallelism $N_G$ and CN number $M$, the larger $r$ becomes, the
longer the per-iteration delay is while the less the required
iteration number becomes as a VN can update its LLR using
information from more VNs. The product of the required iteration
number and the per-iteration delay equals the total decoding delay
to achieve a predetermined error rate performance. Section IV
shows that the NDGSBP algorithm does give improved error rate
performance for the same decoding delay.

\subsection{Basic definitions and notations}
A binary ($N$, $K$) LDPC code $\mathcal{C}$ is a linear block code
whose $M \times N$ parity check matrix $\mathbf{H}=[H_{mn}]$ has
sparse nonzero elements. $\mathbf{H}$ and thus $\mathcal{C}$ can
be viewed as a bipartite graph with $N$ VNs corresponding to the
encoded bits, and $M$ CNs corresponding to the parity-check
functions represented by the rows of ${\mathbf{H}}$.  Given the
above code parameters, the two parameters $r$ and $\ell$ are
related by $\ell\geq\frac{M}{N}\cdot N_G\cdot r$. More
information is needed before an exact relation can be established.
To track the statistical property variations of the
message-passing sequence between VNs and CNs in an iterative
decoding schedule, we also need to know the VN and CN
degree-distribution polynomials $\lambda(x)=\sum^{d_v}_{i=2}
\lambda_ix^{i-1}$ and $\rho(x)=\sum^{d_c}_{j=2}\rho_jx^{j-1}$,
where $\lambda_i$ and $\rho_j$ denote the fraction of all edges
connected to degree-$i$ VNs and degree-$j$ CNs, $d_v$ and $d_c$
denotes the maximum VN and CN degree.

Let $\mathcal{N}$($m$) be the set of variable nodes that
participate in check node $m$ and $\mathcal{M}$($n$) be the set of
check nodes that are connected to variable node $n$ in the code
graph. $\mathcal{N}(m)\backslash n$ is defined as the set
$\mathcal{N}$($m$) with the variable node $n$ excluded while
$\mathcal{M}(n)\backslash m$ is the set $\mathcal{M}$($n$) with
the check node $m$ excluded. Let $L_{n\rightarrow m}$ be the message
sent from VN $n$ to CN $m$ and $L_{m\rightarrow n}$ be the message sent
from CN $m$ to VN $n$.

\subsection{System model and decoding schedule}
Assume a codeword $\mathbf{C}=(c_1,c_2,...,c_{N})$ is
BPSK-modulated and transmitted over an AWGN channel with noise
variance $\sigma^2$. Let $\mathbf{Y}=(y_1,y_2,...,y_{N})$ be the
corresponding received sequence and $L_n$ be the log-likelihood
ratio (LLR) of the variable node $n$ with the initial value given
by $L_n=\frac{2}{\sigma^2}y_n$.

Let $\mathcal{G}_g$ be the $g$th CN group, $1\leq g \leq G$ and
$\mathcal{U}$ be a set of CNs, $l$ as the iteration counter and
$I_{Max}$ as the maximum number of iterations. We can then
describe the NDGSBP algorithm as follows:\\
\\
\noindent\textbf{Initialization}\\
Set $l=1$, $\mathcal{U}=\{x|1\leq x\leq M\}$, and $\mathcal{G}_g=\emptyset$
for $1\leq g\leq G$.

\noindent\textbf{Step 1: Grouping check nodes}\\
Collect $N_G$ elements randomly from the set $\mathcal{U}$ to form
$\mathcal{G}_1$, let $\mathcal{U}$ =
$\mathcal{U}\backslash\mathcal{G}_1$. Collect $N_G-N_G\cdot r$
element randomly from the set $\mathcal{U}$ and $N_G\cdot r$
elements from $\mathcal{G}_{1}$ to create $\mathcal{G}_2$. For
$3\leq g\leq G$, collect $N_G-N_G\cdot r$ element randomly from
the set $\mathcal{U}$ and $N_G\cdot r$ elements from
$\mathcal{G}_{g-1}\backslash\mathcal{G}_{g-2}$ to create
$\mathcal{G}_g$ and let $\mathcal{U}$ =
$\mathcal{U}\backslash\mathcal{G}_g$.

\noindent\textbf{Step 2: Message passing}\\
For $1\leq g\leq G$
\begin{enumerate}
\item[a)] CN update:  $\forall~m \in\mathcal{G}_g, n\in\mathcal{N}(m)$
\begin{equation}
L_{m\rightarrow n}=2\tanh^{-1}\left(\prod_{n'\in
\mathcal{N}(m)\backslash n}\tanh\left(\frac{1}{2}L_{n'\rightarrow
m}\right)\right)
\end{equation}
\item[b)] VN update: $\forall~n \in
\bigcup_{m'\in\mathcal{G}_g}\mathcal{N}(m'), m\in\mathcal{M}(n)$
\begin{equation}
L_{n\rightarrow m}=L_n+\sum_{m'\in \mathcal{M}(n)\backslash m}L_{m'\rightarrow n}
\end{equation}
\end{enumerate}

\noindent\textbf{Step 3: Total LLR computation}\\
$\forall n, 1\leq n\leq N$,
\begin{eqnarray}
L^{total, (l)}_{n}=L_n+\sum_{m'\in N(n)}L_{m'\rightarrow n}
\end{eqnarray}

\noindent\textbf{Step 4: Hard decision and stopping criterion test}\\
\begin{enumerate}
\item[a)] Create
$\textbf{D}^{(l)}=[d^{(l)}_1,d^{(l)}_2,...,d^{(l)}_{N}]$ such
that $d^{(l)}_n=0$ if $L^{total, (l)}_{n}\geq0$ and $d^{(l)}_n=1$ if
$L^{total, (l)}_{n}<0$.\\
\item[b)]  If $\textbf{D}^{(l)}\textbf{H}^T=\textbf{0}$ or
$I_{Max}$ is reached, stop decoding and output $\textbf{D}^{(l)}$
as the decoded codeword. Otherwise, set $l=l+1$ and
$\mathcal{U}=\{x|1\leq x\leq M\}$, go to \textbf{Step 1}.
\end{enumerate}

\section{Convergence Analysis}\label{section:analysis}

\begin{figure}
\begin{center}
\epsfxsize=3.6in \epsffile{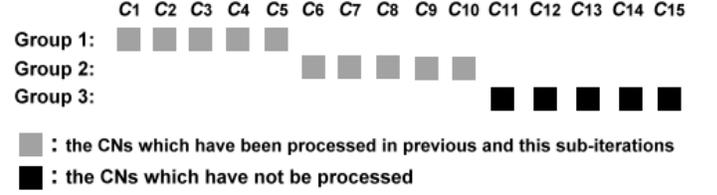}\caption{\label{GS}A
example for GSBP after two sub-iterations.}
\end{center}
\end{figure}

\begin{figure}
\begin{center}
\epsfxsize=3.6in \epsffile{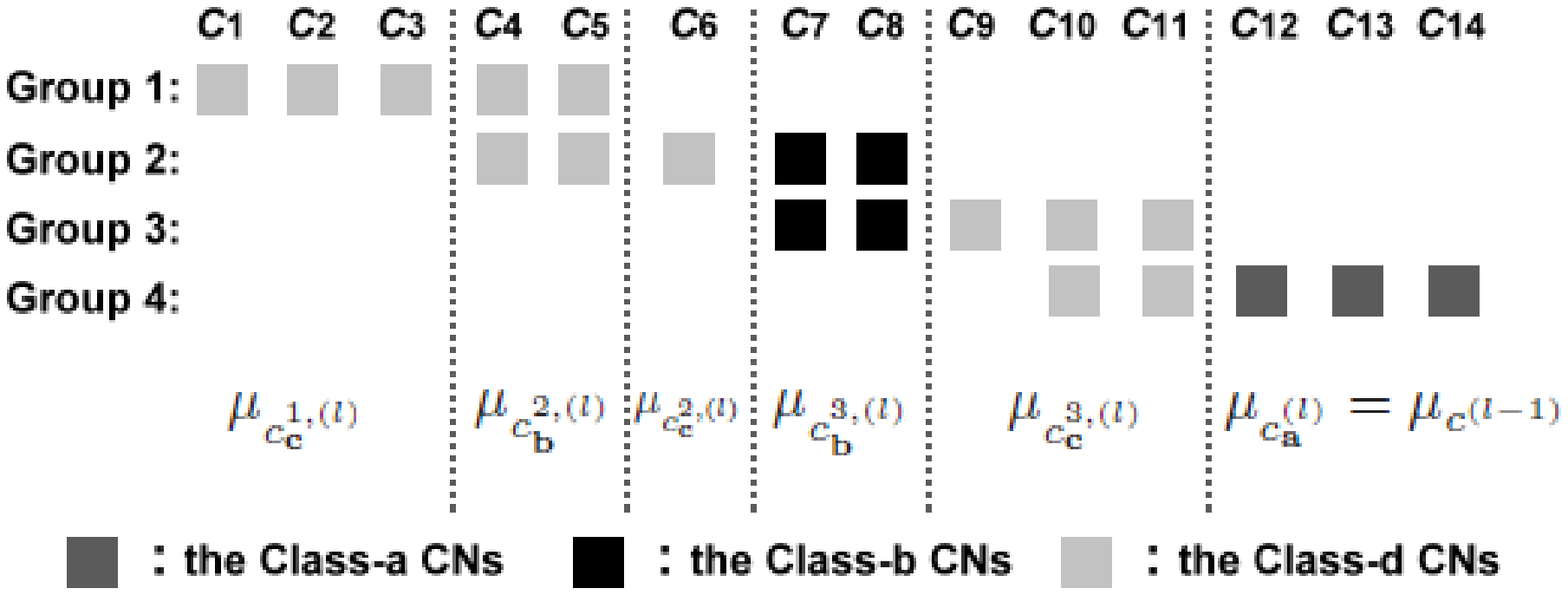}\caption{\label{GOS}A
example for NDGSBP after three sub-iterations when $r<0.5$.}
\end{center}
\end{figure}

\begin{figure}
\begin{center}
\epsfxsize=3.6in \epsffile{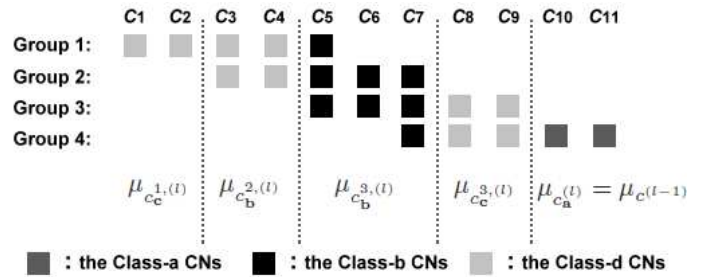}\caption{\label{GOS2}A
example for NDGSBP after three sub-iterations when $0.5\leq r\leq
1$.}
\end{center}
\end{figure}

As can be seen from the above description of the proposed
algorithm, the messages $L_{n\rightarrow m}$ and $L_{m\rightarrow
n}$ are real random variables that depend on the received channel
values $y_n$, the code structure and the decoding schedule. The GA
approach assumes that they can be approximated by Gaussian random
variables. With this approach, we need only to monitor the message
means as the consistency condition holds in our case
\cite{Rich.'01}. We further assume that the all-zero codeword
$\mathbf{C}=(0,0,\dots,0)$, which is mapped into the BPSK
modulated vector $\mathbf{X}=(1,1,\dots,1)$, is transmitted. The
following analysis is based on the ideas of \cite{Chung'01} and
\cite{ZRP'10} with two distinct considerations. First, the
analysis presented in \cite{ZRP'10} deals with vertical GSBP while
we are dealing with horizontal GSBP. Second, the intersection
among groups can be nonempty in our schedule. For GSBP decoding,
we divide CNs into two types, one is updated CNs and the other is
non-updated CNs. As depicted in Fig.\ref{GS}. To analyze the
effect of nonempty intersections, we divide CNs into four classes
in a given, say the $g$th sub-iteration of the $l$th iteration.
Class-\textbf{a} includes the CNs that will be updated at the
$g'$th ($g' > g$) sub-iteration, Class-\textbf{b} includes the CNs
which are also members of the previous $(g-1)$th group,
Class-\textbf{c} contains the CNs which are not members of the
previous $(g-1)$th group and the Class-\textbf{d} are all CNs
exclude Class-\textbf{a} and Class-\textbf{b}. Fig.\ref{GOS} and
Fig.\ref{GOS2} depict the situations after three sub-iterations
for overlapping ratio $r<0.5$ and $0.5\leq r\leq 1$ respectively.

We now track the average values of all updated parameters at the
$l$th iteration for the proposed NDGSBP algorithm. We first define
$\mu_{c^{g,(l)}_{\mathbf{x}}}$ as the mean of the message sent by
a Class-\textbf{x} CN, that is,
$\mu_{c^{g,(l)}_{\mathbf{x}}}=E\{L^{g,(l)}_{m\rightarrow n}\}$,
where $m$ belong to Class-\textbf{x} CNs, $n$ is a VN connecting
to $m$ in the $g$th sub-iteration of the $l$th iteration. We start
with the VN update equation. Consider the degree-$i$ VN $n$ which
is connected to $p$ Class-\textbf{d} CNs, $q$ Class-\textbf{b} CNs
and $i-p-q$ Class-\textbf{a} CNs. For the $g$th sub-iteration of
the $l$th iteration, we have, for $g=1$,
\begin{eqnarray}
\mu_{v^{(l)}_{i,p,q}}=&\mu_{0}+p\mu_{c^{(l)}_{\mathbf{d}}}+q\mu_{c^{g,(l)}_{\mathbf{b}}}\\
\nonumber
&+(i-p-q-1)\mu_{c^{(l)}_{\mathbf{a}}}\\
=&\mu_{0}+p\mu_{c^{(l)}_{\mathbf{d}}}+q\mu_{c^{g,(l)}_{\mathbf{b}}}\\
\nonumber &+(i-p-q-1)\mu_{c^{(l-1)}} \label{eqn:grouporder1}
\end{eqnarray}
where $\mu_{c^{(l)}_{\mathbf{d}}}=\mu_{c^{1,(l)}_{\mathbf{c}}}$
and $\mu_0\triangleq E\{L_n\}=E\{\frac{2y_n}{\sigma^2}\}$ is the
mean of the channel value. For $g>1$, we obtain
\begin{align}
    \nonumber \mu_{c^{(l)}_{\mathbf{d}}}=&\frac{1}{g}\Biggl(\mu_{c^{1,(l)}_{\mathbf{c}}}
    +\mu_{c^{g,(l)}_{\mathbf{c}}}\\
    &+\sum_{g'=2}^{g-1}\biggl(\frac{r}{1-r}\mu_{c^{g',(l)}_{\mathbf{b}}}+\frac{1-2r}{1-r}
    \mu_{c^{g',(l)}_{\mathbf{c}}}\biggr)\Biggr),
\end{align}
for $r<0.5$ and
\begin{eqnarray}
\mu_{c_{\mathbf{d}}^{(l)}}=\frac{1}{g}\left(\mu_{c^{1,(l)}_{\mathbf{c}}}+\mu_{c^{g,(l)}_{\mathbf{c}}}
+\sum_{g'=2}^{g-1}\mu_{c^{g',(l)}_{\mathbf{b}}}\right),
\end{eqnarray}
for $0.5\leq r\leq 1$.

When the CNs in the $g$-th group are processed in $l$-th
iteration, the mean of message for degree-$i$ VNs
$\mu_{v^{(l)}_{i}}$ can be obtained by accumulating all possible
values of $\mu_{v^{(l)}_{i,p,q}}$ with their corresponding
coefficients $\mathrm{\omega}(i,p,q)$:
\begin{equation}
\mu_{v^{(l)}_{i}}=\sum_{p=0}^{i-1}\sum_{q=0}^{i-1-p}\mathrm{\omega}(i,p,q)\cdot\mu_{v^{(l)}_{i,p,q}},
\label{eqn:grouporder2}
\end{equation}
where $\mathrm{\omega}(i,p,q)$ is the proportion of degree-$i$ VNs
which have $p$ neighboring Class-\textbf{d} CNs, $q$ neighboring
Class-\textbf{b} CNs in all degree-$i$ CNs. Thus
$\mathrm{\omega}(i,p,q)$ is given by
\begin{equation}
\mathrm{\omega}(i,p,q)=\left\{\begin{array}{cc}
{i-1\choose p}x^p(1-x)^{i-1-p}, & g=1 \\
{i-1\choose p}{i-1-p\choose q}y^pz^q(1-y-z)^{i-1-p-q}, & g\neq1
\end{array} \right.
\label{eqn:grouporder3}
\end{equation}
where $x$ is the fraction of Class-\textbf{d} CNs for $g=1$, $y$
is the fraction of Class-\textbf{d} CNs and $z$ is the fraction of
Class-\textbf{b} CNs.

Thus
\begin{align}
x=&\frac{1}{G-(G-1)r},\\
y=&\frac{g(1-r)}{G-(G-1)r},\\
z=&\frac{r}{G-(G-1)r}.
\end{align}

From Class-\textbf{c} CNs updating formula, we can obtain
\begin{align}
   E\left\{\tanh\left(\frac{c^{g,(l)}_{\mathbf{c},j}}{2}\right)\right\}=&\left[E\left\{\tanh\left(\frac{v^{(l)}}{2}\right)\right\}\right]^{j-1}.
\end{align}
Under the Gaussian approximation and for $\mu\geq 0$, define
\begin{equation}
    \Phi(\mu)\triangleq1-\frac{1}{\sqrt{4\pi\mu}}\int^\infty_{-\infty}\tanh(\frac{\tau}{2})\exp\left[\frac{-(\tau-\mu)^2}{(4\mu)}\right]d\tau,
\end{equation} and $(13)$ can be rewritten as
\begin{align}
    \mu_{c^{g,(l)}_{\mathbf{c},j}}=&\Phi^{-1}\left(1-\left(1-\sum_{i=2}^{d_v}\lambda_i\Phi\left(\mu_{v^{(l)}_i}\right)\right)^{j-1}\right).
\end{align}
If we average over all CN degree $j$, we have
\begin{align}
    \mu_{c_{\mathbf{c}}^{g,(l)}}=\sum_{j=2}^{d_c}\rho_j\cdot\mu_{c_{\mathbf{c},j}^{g,(l)}}.
\end{align}
The computation of the mean of message send from a
Class-\textbf{b} CN $\mu_{c_{\mathbf{b}}^{g,(l)}}$ is replace
$\mu_{v^{(l)}_i}$ with $\mu_{v'^{(l)}_i}$ in (15) where
$\mu_{v'^{(l)}_i}$ is mean of message send from a previous group
overlapping VN. And $\mu_{v'^{(l)}_i}$ is got by let $p$ at least
$1$ in (8) and (9) for $g\neq1$.

After $l$ iterations, the mean of the message passed from a CN
$\mu_{c^{(l)}}$ is
\begin{align}
    \mu_{c^{(l)}}=\frac{r}{G-(G-1)r}\mu_{c^{G,(l)}_{\mathbf{b}}}+\frac{G-Gr}{G-(G-1)r}\mu_{c^{G,(l)}_{\mathbf{c}}}.
\end{align}
If $\mu_{c^{(l)}}\rightarrow\infty$, the connecting VNs achieve
error free performance.

\section{Numerical Results}\label{section:performance}
\begin{figure}
\begin{center}
\epsfxsize=3.6in \epsffile{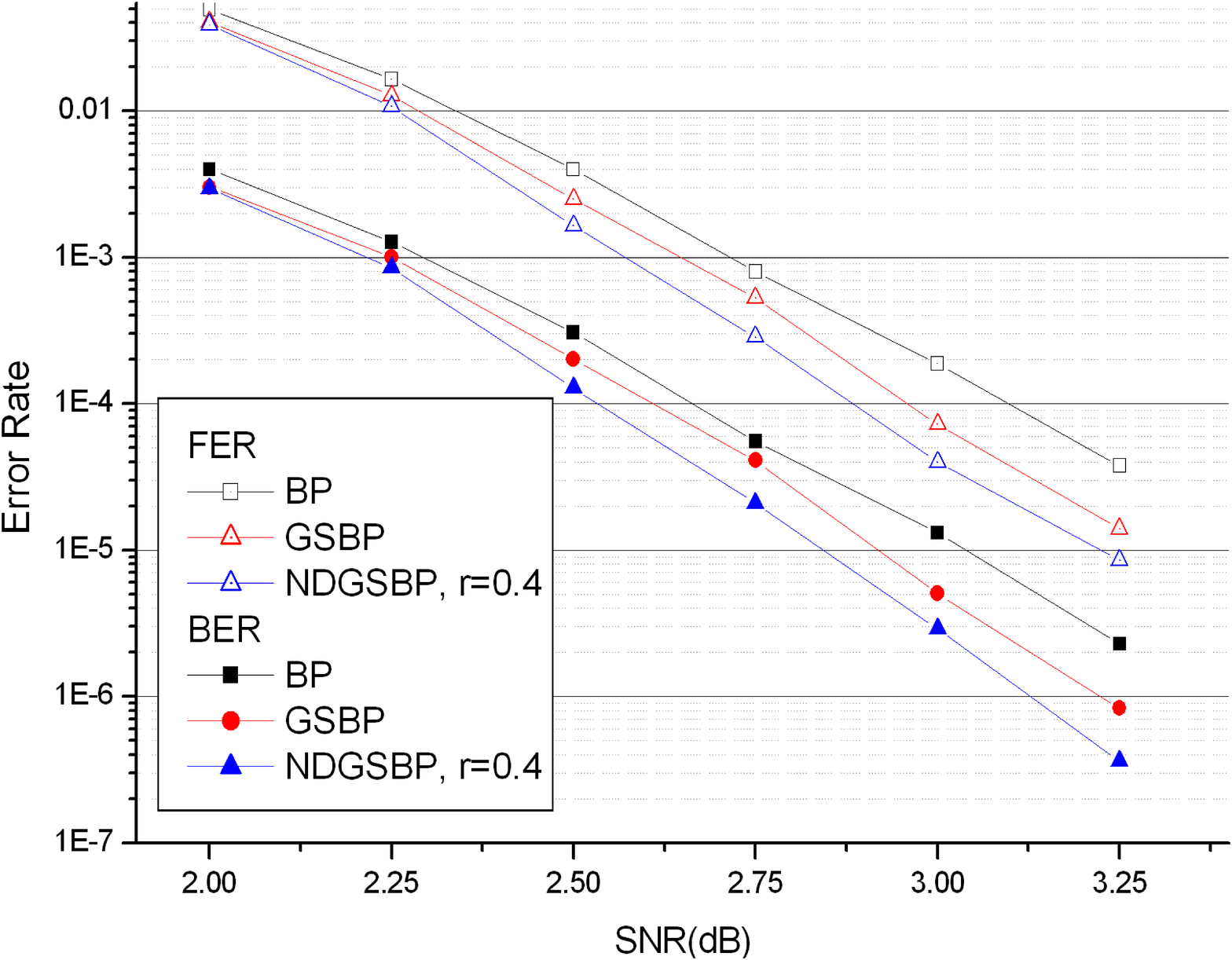}\caption{\label{BER1}BER and
FER performance of Mackay's (504,252) regular LDPC code with
$d_c=6$ and $d_v=3$ using the decoding algorithms: NDGSBP, GSBP
for $G=12$ and standard BP.}
\end{center}
\end{figure}
\begin{figure}
\begin{center}
\epsfxsize=3.6in \epsffile{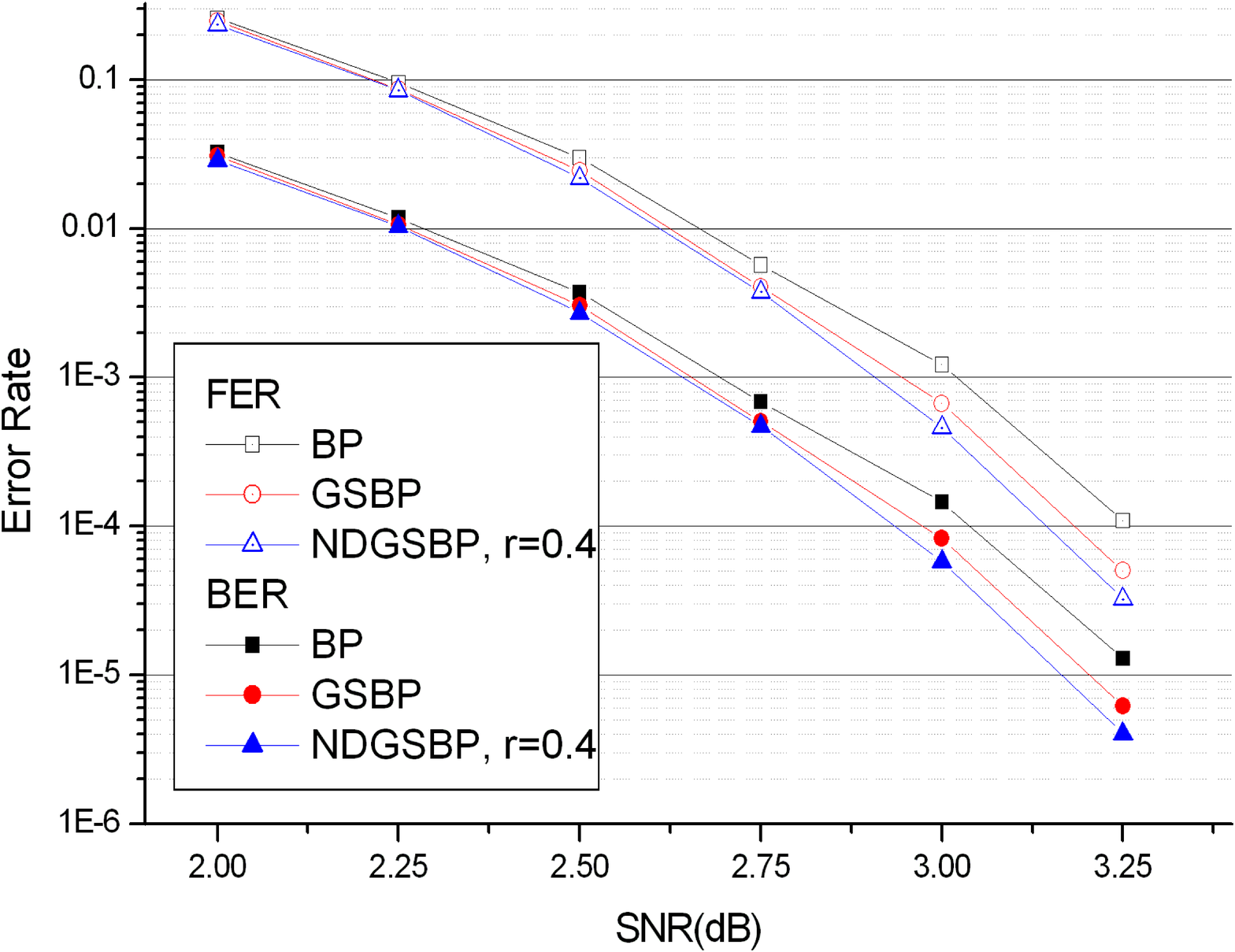}\caption{\label{BER2}BER
and FER performance of Mackay's (816,544) regular LDPC code with
$d_c=6$ and $d_v=4$ using the decoding algorithms: NDGSBP, GSBP
for $G=16$ and standard BP.}
\end{center}
\end{figure}
Fig. \ref{BER1} depicts the FER and BER performance of Mackay's
(504,252) regular LDPC code with $d_c=6$, $d_v=3$ using the
standard BP algorithm, the GSBP algorithm ($G=12$) and the
proposed NDGSBP algorithm ($G=12$, overlapping ratio $r=0.4$). On
the other hand, in Fig. \ref{BER2} we show the FER and BER
performance of Mackay's (816,544) regular LDPC code with $d_c=6$
and $d_v=4$ using the standard BP algorithm, the GSBP algorithm
($G=16$) and the proposed NDGSBP algorithm ($G=16$, overlapping
ratio $r=0.4$).

The simulation results reported in this section assume
$I_{Max}=1000$ for the GSBP and BP algorithms. To have fair
comparison, we assume the system parameter values that result in
the same or similar computation complexity for all algorithms. For
example, to decode the (504,252) LDPC code using the NDGSBP
decoder with $G=12$ and $r=0.4$ imply that $N_G=34$ and it is
allowed to have at most $\frac{m\cdot I_{Max}}{m+(G-1)\times
N_G\cdot r}=\frac{252\cdot 1000}{252+11\cdot 34\cdot 0.4}\approx
627$ decoding iterations.

Fig. \ref{BER1} indicates that at the BER $10^{-5}$, the NDGSBP
decoder is about 0.2dB better than the standard BP decoder, and
achieves about $0.08$dB decoding gain with respect to the the GSBP
decoder. Fig.\ref{BER2} also verify that the performance of the
NDGSBP algorithm is superior to the BP and GSBP algorithm for the
(816,544) LDPC code.

We use the GA approach outlined in Section \ref{section:analysis}
to analyze the performance of the NDGSBP, BP and GSBP decoders.
Given the code rate and degree distribution of LDPC codes, the
thresholds estimated by the GA approach for BP, GSBP and NDGSBP
decoding are the same. In Table \ref{table:GA}, we list the number
of iterations for error free performance at SNR equals threshold.
We examine the NDGSBP performance in decoding two ensemble LDPC
codes using the same overlapping ratio $r=0.4$ but different group
number $G$. The table shows the NDGSBP decoder consistently
outperforms the other two decoders in convergence rate.

\tabcolsep=3pt
\begin{table}
\begin{center} \vspace{2ex}
\caption{\label{table:GA}Number of decoding iterations required to
achieve error-free performance for the BP, GSBP and NDGSBP
($r=0.4$) decoders in a binary-input AWGN channel.}
    \footnotesize{}
    \begin{tabular}{cccccccccccc}
    \hline \hline
    & & & & & GSBP & & & NDGSBP & & &\\
    $d_v$ & $d_c$ & $R$ & $(E_b/N_0)_{\mathrm{GA}}$ & BP & $G=4$ & $12$ & $36$ & $G=4$ & $12$ & $36$ \\
    \hline
    $3$ & $6$ & $1/2$ & $1.163$ & $422$ & $293$ & $262$ & $251$ & $240$ & $208$ & $196$ \\
    \hline
    & & & & & GSBP & & & NDGSBP & & &\\
    $d_v$ & $d_c$ & $R$ & $(E_b/N_0)_{\mathrm{GA}}$ & BP & $G=4$ & $16$ & $34$ & $G=4$ & $16$ & $34$ \\
    \hline
    $4$ & $6$ & $1/3$ & $1.730$ & $632$ & $438$ & $386$ & $376$ & $368$ & $324$ & $317$ \\
    \hline\hline
    \end{tabular}
\end{center}
\end{table}
\section{Conclusions}\label{section:conclusions}
In this paper, we propose a new group shuffled BP decoding
scheduling method to improve the performance of LDPC codes. Our
scheme enhances the connectivity of the code graph by having
overlapped CNs in neighboring CN groups. The enhanced connectivity
allow more each VN (or CN) to obtain related information from more
VNs (or CNs) within a decoding iteration, accelerating the
message-passing rate and thus the convergence speed.

The GA approach is used to track the first-order statistical
information flow of the proposed NDGSBP algorithm. The GA analysis
verifies that the NDGSBP decoder does give faster convergence
performance with respect to that of the GSBP and BP decoders.
Numerical results also demonstrate that, with the same decoding
computation complexity, the new algorithm yields BER and FER
performance better than that of the conventional BP and GSBP
decoders.

In this work, the VN order in grouping is arbitrary and the
non-disjoint parts are randomly selected from the available CNs. A
proper VN ordering and overlapping VN selection that take the code
structure into account will certainly give better performance. The
optimal decoding schedule and parameters ($r$, $\ell$) remain
to be found and some analytic performance metrics may be needed in
our search of the desired solution. \maketitle \linespread{1}


\begin{thebibliography}{16}

%
%

\bibitem{JZhang'05}
J. Zhang and M. Fossorier, ``Shuffled belief propagation
decoding," {\it IEEE Trans. Commun.}, vol. 53, pp. 209-213, Feb.
2005.

\bibitem{DVBS2'06}
A. S¡¦egard, F. Verdier, D. Declercq and P. Urard, ``A DVB-S2
compliant LDPC decoder integrating the horizontal shuffle
schedule," presented in {\it ISPACS¡¦06.}, pp. 1013-1016, Dec.
2006.

\bibitem{Sharon'07}
E. Sharon, S. Litsyn and J. Goldberger, ``Efficient serial
message-passing schedules for LDPC decoding," {\it IEEE Trans.
Inf. Theory}, vol. 53, pp. 4076 - 4091, Nov. 2007.

\bibitem{JZhang'07}
J. Zhang, Y. Wang, M. P. C. Fossorier and J. S. Yedidia,
``Iterative decoding with replicas," {\it IEEE Trans. Inf.
Theory}, vol. 53, No. 5, pp. 1644-1663, May 2007.

\bibitem{ccylucky^n'09}
C.-Y. Chang, Y.-L. Chen, C.-M. Lee, and Y. T. Su, ``New group
shuffled BP decoding algorithms for LDPC codes," presented in {\it
IEEE Int. Symp. on Inform. Theory 2009}, pp. 1664-1668, Jun. 2009.

\bibitem{Yang'10}
Y. Yang, J.-Z. Huang, S. Tong and X.-M. Wang, ``Replica
horizontal-hhuffled iterative decoding of low-density parity-check
codes," in {\it The Journal of China Universities of Posts and
Telecommunications}, vol. 13, pp.32-40, Jun. 2010.

\bibitem{Rich.'01}
T. J. Richardson and R. L. Urbanke, ``The capacity of low-density
parity-check codes under message-passing decoding," {\it IEEE
Trans. Inf. Theory}, vol. 47, No. 2, pp. 599-617, Feb. 2001.

\bibitem{Chung'01}
S.-Y. Chung and T. J. Richardson, ``Analysis of sum-product
decoding of low-density parity-check codes using a Gaussian
approximation," {\it IEEE Trans. Inf. Theory,} vol. 47, No. 2, pp.
657-670, Feb. 2001.

\bibitem{Brink'04}
S. T. Brink and G. Kramer, ``Design of low-density parity-check
codes for modulation and detection," {\it IEEE Trans. commun.},
vol. 52, No. 4 pp. 670-678, Apr. 2004.

\bibitem{Sharon'06}
E. Sharon and A. Ashikhmin, ``Analysis of low-density parity-check
codes based on EXIT functions," {\it IEEE Trans. commun.}, vol.
54, No. 8, pp. 1407-1414, Aug. 2006.

\bibitem{ZRP'10}
Z. Song, R. Yu, and P. Ma, ``Gaussian approximation for LDPC codes
under group shuffled belief propagation decoding," presented in
{\it IEEE Int. Conference on Wireless Commun., Networking and
Mobile Computing 2010}, pp. 1-4, sep. 2010.

\end{thebibliography}
\end{document}